\documentclass[arps,twocolumn,showpacs]{revtex4}

\usepackage{bm}
\usepackage{graphicx}
\usepackage{subfigure}
\input epsf

\newcommand\lsim{\mathrel{\rlap{\lower4pt\hbox{\hskip1pt$\sim$}}
        \raise1pt\hbox{$<$}}}
\newcommand\gsim{\mathrel{\rlap{\lower4pt\hbox{\hskip1pt$\sim$}}
        \raise1pt\hbox{$>$}}}

\begin{document}
\title{The Evolution of Bias - Generalized}

\author{Lam Hui$^{1,2}$ and Kyle P. Parfrey$^{1,3}$}

\affiliation{
$^{1}$Institute for Strings, Cosmology and Astroparticle Physics (ISCAP)\\
$^{2}$Department of Physics, Columbia University, New York, NY 10027, U.S.A.\\
$^{3}$Department of Astronomy, Columbia University, New York, NY 10027, U.S.A.\\
lhui@astro.columbia.edu, kyle@astro.columbia.edu
}
\date{\today}

\begin{abstract}
Fry (1996) showed that galaxy bias has the tendency to evolve towards unity, 
i.e. in the long run, the galaxy distribution tends to trace that of matter.
Generalizing slightly Fry's reasoning,
we show that his conclusion remains valid in theories of modified
gravity (or equivalently, complex clustered dark energy). This is not surprising:
as long as both galaxies and matter are subject to the same force, dynamics would
drive them towards tracing each other. This holds, for instance,
in theories where both galaxies and matter move on geodesics.
This relaxation of bias towards unity is tempered
by cosmic acceleration, however: the bias tends towards unity but
does not quite make it, unless the formation bias were close to unity.
Our argument is extended in a straightforward 
manner to the case of a stochastic or nonlinear bias.
An important corollary is that dynamical evolution could imprint a scale dependence
on the large scale galaxy bias. This is especially pronounced if non-standard gravity
introduces new scales to the problem: the bias at different scales relaxes at different
rates, the larger scales generally more slowly and retaining a longer memory of the initial bias.
A consistency test of the current (general relativity $+$
uniform dark energy) paradigm is therefore to look for departure from a scale independent
bias on large scales. 
A simple way is to measure the relative bias
of different populations of galaxies which are at different stages of bias relaxation.
Lastly, we comment on the possibility of directly testing the Poisson equation
on cosmological scales, as opposed to indirectly through the growth factor.
\end{abstract}


%%98.70.Vc is for Background radiations

\maketitle

\section{Introduction}
\label{intro}

In a classic paper, Fry \cite{jimfry} showed that galaxies, once formed,
would evolve to eventually trace mass. This result was subsequently extended
to the case of a stochastic bias by Tegmark and Peebles \cite{max}.
Similar results were found in numerical simulations \cite{simulations}. The linear
bias paradigm has been applied to a variety of cosmological models, including $\Lambda$CDM
and quintessence \cite{spyros}.   
Both \cite{jimfry} and \cite{max} assumed
Newtonian/Einsteinian gravity from the start. One of our principal aims is
to show that relaxing this assumption does not alter the basic conclusions.
In the process, we will uncover an interesting, unanticipated corollary.

The basic reasoning is simple: both galaxies and (dark) matter can
be regarded as test particles in some gravitational field; since their
dynamics are identical, they are expected to trace each other given
sufficient time. This holds regardless of the origin of the gravitational field.
Departures from Newtonian/Einsteinian gravity on large scales have
been suggested as a possible explanation for the observed cosmic acceleration. 
Most of these theories of modified gravity, for instance DGP \cite{DGP} or
scalar-tensor theories \cite{carroll}, are metric theories, and matter (or galaxy) 
particles move on geodesics defined by the metric. This is true as long as
matter and galaxies are minimally coupled to gravity in the relevant frame
\cite{frame}. The statement that matter and galaxies should evolve to trace
each other does not rely on how the metric comes about. All that matters
is that they have the same dynamics under the metric.
More generally, as we will see, our conclusions on bias evolution remain
valid as long as both galaxies and matter are subject to the same force.

Note that the same conclusion holds if, instead of modifying gravity,
one introduces additional sources of fluctuations, for instance clustered
dark energy. 
The presence of dark energy perturbations modifies the relation
between matter perturbations and the metric, or the gravitational potential
(just as modifying gravity does). But once this gravitational potential
is given, both galaxies and matter particles respond to it in the same way,
therefore guaranteeing eventual evolution towards a trivial biasing relation between them.
This is consistent with the fact that modifying gravity can be viewed
as some complicated form of dark energy - 
no surprise given that one can always shift terms from the left hand side
of the Einstein equation to the right \cite{equiv}.
Modifications of gravity are indistinguishable from complex dark energy, at least at
the linear level,
unless constraints are put on the nature of dark energy, such as 
spatial homogeneity or zero coupling to matter (e.g. \cite{wiley,bhuv} and references therein).

Why are we interested in proving these general but in a sense rather obvious
statements about bias evolution? 
First, as we will see, there is an interesting 
and probably not so obvious corollary: that bias evolution can imprint
a scale dependence even if the initial bias were scale independent.
The effect is especially large if one's non-standard theory of gravity/dark energy 
introduces new scales to the problem, making the growth factor scale dependent. 
A promising way to test the
current (general relativity $+$ uniform dark energy) paradigm is therefore
to look for a scale dependent bias. 
Our second (and original) motivation is to address the issue
of directly testing the Poisson equation. Most observations,
such as gravitational lensing and the microwave background, tell us
the metric fluctuations or gravitational potential, the left hand side of
the Poisson equation. There are very few observations that tell us directly
what sources the gravitational potential, the right hand side of the Poisson equation.
The galaxy distribution provides one such handle, but it is complicated
by the uncertain galaxy bias. One might hope that the tendency of the galaxy bias
to relax towards unity could help bridge the gap between
galaxies and mass. However, as we will see, in an accelerating universe, the bias
relaxation process is quite slow. Modifications of gravity that predict
cosmic acceleration tend to slow it further.
The slow relaxation process makes it difficult to argue that galaxies that formed
at high redshifts tend to have a trivial bias today, unless they formed with a bias close to unity. On the other hand, as we will see, the slow bias relaxation
is a boon to the scale dependence test (motivation number one).

Two clarifying remarks are in order. First, throughout this paper, the galaxy bias
is defined with respect to the matter fluctuations.
It is important to bear this in mind since, in models that allow clustered dark energy, 
there are different kinds of galaxy bias: with respect to matter, dark energy
or the sum.
From the point of view of dynamical evolution, the bias with respect to matter
is the natural one to consider.
Secondly, an important assumption in \cite{jimfry} as well as here is
that the galaxy number is conserved, 
just as mass is. In practice, galaxies are of course created (they form)
and destroyed (they merge). 
%Nonetheless, it is
%conceivable that one can isolate the bias evolution effect described here by
%selecting galaxies that, as a population, formed a long time ago, and
%have not suffered significantly from mergers since.
%This is quite tricky, as will be further discussed below.
%In any case, bias evolution is certainly part of the story for
%any given sample of galaxies: a prediction for bias needs to fold in
%the dynamical evolution in addition to formation and destruction.
A complete theory of galaxy bias needs to fold in these processes.
We will have more to say about this in \S \ref{discuss}.
Dynamical evolution is in any case an important part of the story.
%, and
%our calculations should apply without modification to certain populations of
%galaxies. Exactly what populations will be discussed further below.

This paper is organized as follows. 
The basic equations are reviewed in
\S \ref{basics}. The implications for the linear galaxy bias are
worked out in \S \ref{linear}. The extension to a (linear) stochastic bias is
presented in \S \ref{stochastic}, and the extension to a nonlinear
bias is given in \S \ref{nonlinear}. 
The interesting corollary of a possible scale dependent imprint on
the galaxy bias is derived and discussed at the end of both
\S \ref{linear} and \ref{stochastic}.
In \S \ref{discuss}, we conclude with some remarks about 
testing theories of modified gravity or complex dark energy.

\section{Basics}
\label{basics}

According to the current structure formation paradigm, 
the fundamental equations governing the dynamics of sub-Hubble
matter fluctuations are \cite{peebles}:
\begin{eqnarray}
\label{massconserve}
\delta'_m = - {\bf \nabla} \cdot [(1+\delta_m) {\bf v_m}] \\
\label{momconserve}
{\bf v'_m} + ({\bf v_m} \cdot {\bf \nabla})\, {\bf v_m} + {a' \over a} {\bf v_m} = - {\bf \nabla} \phi \\
\label{poisson}
\nabla^2 \phi = 4 \pi G \bar\rho_m a^2 \delta_m
\end{eqnarray}
where $\delta_m \equiv (\rho_m - \bar\rho_m)/\bar\rho_m$ is the
matter overdensity, with $\rho_m$ being the matter mass density
and $\bar\rho_m$ its spatial average, ${\bf v_m}$ is the matter 
peculiar velocity,
$a$ is the scale factor, $\phi$ is the gravitational potential or metric perturbation,
$G$ is Newton's constant, $'$ denotes a derivative with respect
to conformal time $\eta$, and ${\bf \nabla}$ denotes a derivative with respect to
comoving position ${\bf x}$.

Equations (\ref{massconserve}) and (\ref{momconserve}) express
respectively mass and momentum conservation.
Equation (\ref{poisson}) is the Poisson equation, which tells us
how matter fluctuations source the gravitational potential according
to Newtonian/Einsteinian gravity.

Galaxies (or more precisely, some population thereof), 
once formed, obey the following equations on sub-Hubble scales:
\begin{eqnarray}
\label{numconserve}
\delta'_g = - {\bf \nabla} \cdot [(1+\delta_g) {\bf v_g}] \\
\label{momgconserve}
{\bf v'_g} + ({\bf v_g} \cdot {\bf \nabla})\, {\bf v_g} + {a' \over a} {\bf v_g} = - {\bf \nabla} \phi
\end{eqnarray}
where $\delta_g \equiv (n_g - \bar n_g)/\bar n_g$ is the 
galaxy overdensity, with $n_g$ being the galaxy number density
and $\bar n_g$ its spatial average, and ${\bf v_g}$ is the galaxy
peculiar velocity.
Equation (\ref{numconserve}) crucially assumes number conservation, a subject to
which we will return in \S \ref{discuss}. Equation (\ref{momgconserve}) is
momentum conservation, or the equation of motion, for the galaxies.
It is worth noting that we do not assume ${\bf v_m} = {\bf v_g}$, as is often done.
This is in part because we want to make explicit the importance of subjecting the
galaxies and matter to the same force. If they are not subject
to the same force, ${\bf v_m}$ and ${\bf v_g}$ would diverge from each other;
if they are, the two velocities tend to converge, as we will see.

Equations (\ref{massconserve}) to (\ref{momgconserve}) constitute the starting
point for \cite{jimfry,max} on the evolution of bias. 
Our key observation is that Eq. (\ref{poisson}) is
unnecessary, at least on a qualitative level.

This is interesting in light of the recent
interest in theories of modified gravity. By and large, these theories
respect mass and momentum conservation; particles (matter or galaxy)
move on geodesics. The only equation that is modified in these theories
is therefore the Poisson equation (\ref{poisson}). 
Similarly, in theories where dark energy has non-trivial clustering
behavior, dark energy can act as an additional source for gravitational perturbations,
and therefore the simple form of the Poisson equation given in (\ref{poisson}) is again modified. 
%Indeed, modifying gravity and complex, non-uniform dark energy are but two sides
%of the same coin \cite{equiv}.

Henceforth, our derivation relies on Eqs. (\ref{massconserve}), (\ref{momconserve}),
(\ref{numconserve}) and (\ref{momgconserve}), but not on (\ref{poisson}).
In other words, our only assumptions are: that mass and number are conserved, and that
matter and galaxies are subject to the same force. 

\section{Linear Bias}
\label{linear}

Keeping only terms linear in fluctuations, one can see 
from Eqs. (\ref{momconserve}) and (\ref{momgconserve}) that
\begin{eqnarray}
\label{vdiff}
({\bf v_g} - {\bf v_m})' + {a' \over a} ({\bf v_g} - {\bf v_m}) = 0
\end{eqnarray}
which implies
\begin{eqnarray}
\label{vdiff1}
({\bf v_g} - {\bf v_m}) = {\bf u_0} / a
\end{eqnarray}
where ${\bf u_0}$ is independent of time but dependent on position. 
Note that in deriving this result, we do not need the right hand sides of
Eqs. (\ref{momconserve}) and (\ref{momgconserve}) to be the gradient of some potential;
all we need is that they are identical i.e. matter and galaxies are subject to the same force.
Equation (\ref{vdiff1}) tells us that if ${\bf v_g} = {\bf v_m}$ initially, it will
remain so; any initial velocity bias, should it exist, decays away.
This holds in the context of linear perturbation theory (i.e. large scales)
\cite{dynfric}.
In fact, \cite{jimfry,max} set the galaxy and matter velocities to be 
equal from the outset.

Similarly, linearizing Eqs. (\ref{massconserve}) and (\ref{numconserve}) and taking the
difference, we obtain
\begin{eqnarray}
\label{deltadiff}
(\delta_g - \delta_m)' = - {\bf \nabla} \cdot ({\bf v_g - v_m}) = - {\bf \nabla} \cdot {\bf u_0} / a \;.
\end{eqnarray}
This can be integrated to give
\begin{eqnarray}
\label{deltadiff1}
\delta_g - \delta_m = - {\bf \nabla} \cdot {\bf u_0} \int {da \over a^3 H} + \Delta_0
\end{eqnarray}
where $\Delta_0$ is independent of time but spatially dependent, and $H$ is the Hubble parameter.
The integral over $a$ generally gives a decaying term in contrast with
the constant $\Delta_0$ term, and so to leading order, the
galaxy bias evolves as:
\begin{eqnarray}
\label{linearbias}
b(a) \equiv {\delta_g (a)\over \delta_m(a)} = 1 + {\Delta_0 \over \delta_m(a)}
\end{eqnarray}
where we have inserted the argument $a$ in each quantity that
is time dependent (while space dependence is kept implicit). 
The quantity $\Delta_0$ is independent of time, and can be alternatively expressed
as
\begin{eqnarray}
\label{Delta0alt}
\Delta_0 = (b_0 - 1) \delta_m (a_0)
\end{eqnarray}
where $b_0$ is the initial bias at some early time corresponding to 
scale factor $a_0$.
From Eq. (\ref{linearbias}), we can see that
as long as the matter fluctuation $\delta_m$ grows, 
the galaxy bias $b$ evolves towards unity.
Exactly how rapidly it evolves depends on the growth rate of $\delta_m$, which cannot be
determined unless the modification to Eq. (\ref{poisson}), or lack thereof, is specified.
For small fluctuations, it is not unreasonable to expect that 
$\delta_m$ grows roughly on the Hubble time scale. 

There is an interesting corollary that follows from Eq. (\ref{linearbias}). 
Suppose the galaxy bias is initially scale independent.
This means $\delta_g (a_0, {\bf x}) = b_0 \delta_m (a_0, {\bf x})$,
where $b_0$ is the initial bias which is independent of position,
and $a_0$ denotes the scale factor at some early time. Equations
(\ref{linearbias}) and (\ref{Delta0alt}) then tell us for $a \ge a_0$:
\begin{eqnarray}
\label{linearbias2}
b(a, {\bf x}) = 1 + (b_0 - 1) {\delta_m (a_0, {\bf x}) \over \delta_m (a, {\bf x})}
\end{eqnarray}
which means $b$ is independent of position, or is scale independent,
only if $\delta_m (a, {\bf x})$ preserves exactly the initial spatial
dependence encoded in $\delta_m (a_0, {\bf x})$. We know this holds
in general relativity (plus uniform dark energy) 
at the level of linear theory in the sub-Hubble limit.
One can define the bias $b$ in Fourier space as well, in which
case Eq. (\ref{linearbias2}) takes the form:
\begin{eqnarray}
\label{linearbias3}
b(a, {\bf k}) = 1 + (b_0 - 1) {\delta_m (a_0, {\bf k}) \over \delta_m (a, {\bf k})} \;.
\end{eqnarray}
In the standard paradigm of general relativity plus uniform dark energy,
different sub-Hubble Fourier modes of $\delta_m$ grow with the same ($k$ independent) 
growth factor, hence an initially
scale independent bias ($b_0$) will remain scale independent ($b$).
Note that according to the standard paradigm, the initial
linear bias is expected to be scale independent \cite{mowhite,shethtormen, scherrerweinberg}.

\begin{figure}[tb]
\centerline{\epsfxsize=9cm\epsffile{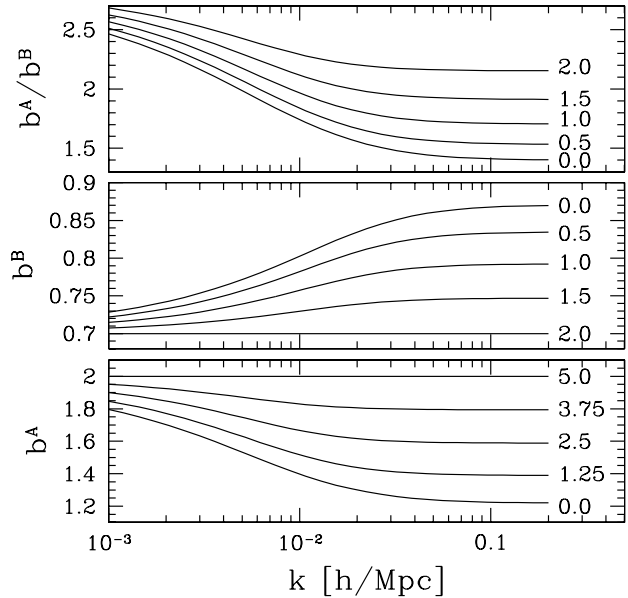}}
\caption{The evolution of the linear bias and relative bias as a function of wavenumber $k$.
In the bottom panel is $b^A$, the linear bias of galaxies that formed at redshift $5$ with an initial
scale independent bias of $b_0 = 2$. In the middle panel is $b^B$, the linear bias of galaxies that 
formed at redshift $2$ with
an initial scale independent bias of $b_0 = 0.7$. The top panel shows the ratio of the two.
The number labeling each curve is the corresponding redshift.
This makes use of the Yukawa modification to the Poisson equation: Eq. (\ref{yukawa}).
}
\label{bevolveB.yukawa}
\end{figure}

However, it is entirely
possible that modifications to the Poisson equation (\ref{poisson}) 
take a scale dependent form, which would make the perturbation growth scale
dependent.
General relativity itself introduces such corrections, but
they are generally suppressed by $(aH/k)^2$ and are therefore very small except
close to the Hubble scale where cosmic variance is large \cite{amin}.
A less trivial example is a Yukawa-like modification of the Poisson equation:
$\nabla^2 \rightarrow \nabla^2 - \mu^2$, where $1/\mu$ defines some new scale \cite{white,verde}.
Scalar-tensor theories introduce modifications of this kind where
$\mu^2$ is related to the second derivative of the scalar potential \cite{wagoner}, in addition
to the better known effect of giving $G$ a time dependence \cite{will,lobo,boisseau}.
In these cases, the linear growth factor $\delta_m (a, {\bf k})/\delta_m (a_0, {\bf k})$ will
be $k$ dependent. As a concrete example, we show in the bottom two panels of
Fig. \ref{bevolveB.yukawa} the linear bias as a function of redshift and wavenumber $k$ for
two different populations of galaxies, with the following modification to the Poisson equation:
\begin{eqnarray}
\label{yukawa}
(\nabla^2 - a^2 m^2)\phi = 4\pi G \bar\rho a^2 \delta_m
\end{eqnarray}
where $m$ is a constant and
$a^2 m^2$ plays the role of $\mu^2$. We use $\mu \propto a$ such that
the Yukawa correction becomes small in the early universe \cite{verde},
although some other time dependence is possible. There is also the question
of how such a scalar-tensor theory can be made consistent with solar system constraints;
this depends on the full Lagrangian of the theory - the kinetic term, the potential and
the coupling to curvature can always be chosen to satisfy such constraints
\cite{chameleon}.
In Fig. \ref{bevolveB.yukawa}, we adopt $m = 0.05$ h/Mpc
which appears to be consistent with current observations \cite{white,verde}.
A flat cosmology of $\Omega_m = 0.27$ and $\Omega_\Lambda = 0.73$ is assumed in
computing the background expansion rate.

From Eq. (\ref{yukawa}), one can see that the high $k$ modes ($k/a \gg m$) are insensitive
to the Yukawa correction. Hence, the high $k$ end of Fig. \ref{bevolveB.yukawa} shows
a bias evolution that is consistent with Einsteinian/Newtonian gravity.
It is worth stressing that while the bias {\it wants} to evolve towards unity, it does not
quite make it to unity.
This is the result of cosmic acceleration: perturbation growth is much suppressed once the
universe starts to accelerate, and the bias relaxation process is stalled.
In fact, even if we continue the evolution into the future,
the bias will not become very close to unity, unless its initial 
value were close to unity to begin with.
%This is the result of cosmic acceleration: growth is much suppressed once the
%universe starts accelerating; for our adopted parameters, 
According to general relativity, 
$\delta_m (a=\infty)/\delta_m (a=1) \sim 1.36$ for our adopted parameters, implying
bias relaxation via Eq. (\ref{linearbias3}) is rather limited in the future.
%%See Code/Test_GR/growfk.dat and D.ps.

If general relativity were the whole story,
the curves in Fig. \ref{bevolveB.yukawa} would have been flat
(aside from very small $(aH/k)^2$ corrections mentioned earlier).
Instead, we see that the Yukawa correction further suppresses the growth of perturbations at
low $k$'s (compared to high $k$'s), which significantly slows their evolution towards a bias of unity.
This is qualitatively what one expects from theories of modified gravity
which are constructed to yield cosmic acceleration: such theories tend to make gravity
weaker on large scales.
In other words, the lower-$k$ modes retain a longer memory of the initial bias, while
the higher-$k$ modes catch up more efficiently (but not completely) 
with the matter perturbations. The result is a scale dependent bias, 
as shown in Fig. \ref{bevolveB.yukawa}. 
Note that, in principle, this scale dependence is temporary -- if
the bias does eventually reach unity in the far future, it becomes scale independent
once again. In practice, in an accelerating universe, the bias relaxation process
is so slow that this does not actually occur, unless the initial bias were quite
close to unity. 
This is a useful feature from the perspective of preserving, and
potentially observing, the dynamically generated scale dependent bias.

We note in passing that the rather popular $f(R)$ theory is 
a scalar-tensor theory, and is therefore
expected to exhibit a scale dependent bias qualitatively
similar to that shown in Fig. \ref{bevolveB.yukawa}.
It appears, however, that the particular $f(R)$ theories
usually considered in the literature have an effective mass $m$ (or $\mu$) that
is rather small \cite{carroll}, which makes for a weak scale dependence.
There is no fundamental reason, though, to focus only on these particular
realizations of a scalar-tensor theory.

\begin{figure}[tb]
\centerline{\epsfxsize=9cm\epsffile{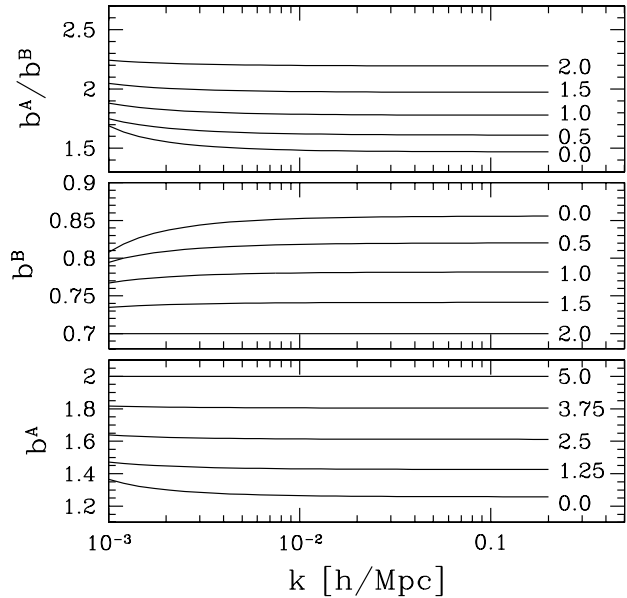}}
\caption{Analog of Fig. \ref{bevolveB.yukawa} except that 
a DGP motivated modification of growth rate is used: Eq. (\ref{DDGP}).
Note that this does not account for the scale dependence arising
from the transition from the scalar-tensor regime to the general relativity
regime, the so called $r_*$ effect.
}
\label{bevolveB.DGP}
\end{figure}

Another example of a modified gravity theory is DGP \cite{DGP}, which is expected
to introduce scale dependence to the growth factor as well.
There are two sources of scale dependence. One is that DGP is expected to
give corrections to the growth factor of the order of $aH_0/k$, arising 
from its peculiar nonlocal character from the 4D perspective (Scoccimarro, private
communication). The other is that DGP introduces a special scale, often referred to
as $r_*$, where gravity makes a transition from the general relativity regime to a scalar-tensor regime
\cite{rstarcomment}. 
One expects the structure growth to have a feature around $r_*$.
Its precise value is however uncertain, because it depends on
the exact nature and characteristics of the fluctuations. None of the existing
calculations of structure growth in DGP gravity address growth in this transition
regime \cite{DGPgrowth}. The general expectation is that $r_*$ is close to the
nonlinear scale, but its actual value could well be larger.
As an illustration, in Fig. \ref{bevolveB.DGP}, we sidestep the $r_*$ issue,
and simply modulate the DGP growth by an $aH_0/k$ correction (see also \cite{amin}):
\begin{eqnarray}
\label{DDGP}
\delta_m^{\rm DGP} (a, {\bf k}) = \delta_m^{\rm Poisson} (a, {\bf k}) (1 - aH_0/k)
\end{eqnarray}
where $\delta_m^{\rm Poisson}$ denotes the matter fluctuation according to 
the Poisson equation, with a background expansion chosen to match
the expected DGP growth on small scales (but larger than $r_*$) \cite{wileynote}.
Here, the scale dependence of the bias is much weaker than in the Yukawa example, though
one must remember that this ignores the $r_*$ effect.
As mentioned before, general relativity itself introduces corrections
of this sort, but they come in at quadratic order $(aH/k)^2$ and are even weaker.

Lastly, instead of modifying the Poisson equation (\ref{poisson}) on the left
hand side, one could also modify it on the right hand side 
by having dark energy that clusters. This would introduce new scales
to the problem, such as the Jeans scale, and one expects the growth of perturbations,
and therefore the bias relaxation, to become scale dependent.

%It is important to emphasize that any dynamically induced scale dependence to the bias is
%temporary. As can be seen from Eqs. (\ref{linearbias2}) or (\ref{linearbias3}),
%the bias eventually relaxes to unity. Therefore, the scale dependence
%exists only in so far as the bias retains some memory of its initial value.
%The dynamically generated scale dependence enters through the mismatch between
%the initial fluctuations ($\propto \delta_m (a_0, {\bf k})$) and the growing
%fluctuations ($\propto \delta_m (a, {\bf k})$) i.e. that their ratio could be
%$k$ dependent. Once the memory of the initial fluctuations
%fades away, the bias becomes unity and therefore scale independent.

We should add that in theories where gravity is modified, or clustered dark
energy is introduced, it is quite possible that the galaxies are {\it born} with
a scale dependent bias on large scales, contrary to expectations according to the standard
excursion set theory of halos \cite{mowhite,shethtormen}.
It would be very surprising if any such initial scale dependence conspires to cancel 
the scale dependence that develops dynamically.
Checking for departure from a scale independent bias on large scales therefore
constitutes an interesting test of (scale dependent) modifications to the Poisson equation 
(\ref{poisson}).
One way to perform this test is to compare the galaxy power spectrum
and the mass power spectrum from lensing
(the latter requires the Poisson equation to convert from metric to
mass fluctuations).
A simpler test is to compare the large scale bias of different
populations of galaxies, which are at different stages of dynamical bias relaxation. For instance, labeling the two populations of galaxies by
$A$ and $B$, their relative bias is
\begin{eqnarray}
\label{relativebias}
{b^A (a, {\bf k}) \over b^B (a, {\bf k})}
= {1 + (b_0^A-1) {\delta_m (a_0^A, {\bf k})/ \delta_m (a, {\bf k})} \over
1 + (b_0^B - 1) {\delta_m (a_0^B, {\bf k})/ \delta_m (a, {\bf k})}}
\end{eqnarray}
for $a \ge a_0$. A ratio of this kind can be obtained from the ratio
of the observed power spectra of $A$ and $B$. The top panels of 
Fig. \ref{bevolveB.yukawa} and \ref{bevolveB.DGP} illustrate
this ratio for two particular populations of galaxies.
It is worth noting that the best chance for observing a scale dependent relative bias is to
compare overbias ($> 1$) galaxies with underbias ($< 1$) galaxies. 
This is because modified gravity theories tend to predict
a higher bias for the overbiased galaxies, and a lower bias for
the underbiased galaxies. This effect gets stronger at lower $k$'s;
the scale dependence is most visible in their relative bias.
So far, there is no observational evidence for
departure from scale independence on large (i.e. linear) scales \cite{scaleindep}.

\section{(Linear) Stochastic Bias}
\label{stochastic}

Having a stochastic bias means one introduces an additional parameter
into the problem, following Dekel \& Lahav \cite{dekellahav}.  In addition to $b$, defined by,
\begin{eqnarray}
\label{bdefine}
b^2 \equiv \langle \delta_g^2 \rangle / \langle \delta_m^2 \rangle
\end{eqnarray}
one introduces $r$, defined by
\begin{eqnarray}
\label{rdefine}
br \equiv \langle \delta_g \delta_m \rangle / \langle \delta_m^2 \rangle
\end{eqnarray}
Here, one can think of the ensemble averages in several different ways.
For instance, $\langle \delta_g \delta_m \rangle$ can be thought of as 
the ensemble average of the product of $\delta_g$ and $\delta_m$, each
smoothed on some given scale $R$. It can also be thought of as 
the two point correlation function of $\delta_g$ and $\delta_m$ at
some given comoving separation $\Delta x$. Another alternative is to
think of it as the power spectrum at a wavenumber $k$: 
$\propto [\langle \delta_g ({\bf k}) \delta_m^* ({\bf k}) \rangle + 
\langle \delta_g^* ({\bf k}) \delta_m ({\bf k}) \rangle]/2$.
Depending therefore on the definition, $b$ and $r$ are in general functions
of $R$, $\Delta x$ or $k$ respectively.
It is worth noting that cross-correlation measurements of galaxies
suggest $r$ is consistent with unity on large scales \cite{scaleindep}.

Squaring Eq. (\ref{deltadiff1}),
and keeping only the leading contribution to the ensemble average,
we have
\begin{eqnarray}
\label{deltadiff1stoc}
b^2 - 2br + 1 = \langle \Delta_0^2 \rangle / \langle \delta_m^2 \rangle
\end{eqnarray}
where $\langle \Delta_0^2 \rangle$ is time independent, but $\langle \delta_m^2 \rangle$ is not.
Alternatively, multiplying Eq. (\ref{deltadiff1}) by $\delta_m$, and keeping
only the leading contribution:
\begin{eqnarray}
\label{deltadiff2stoc}
br - 1 = \langle \Delta_0 \delta_m \rangle / \langle \delta_m^2 \rangle
\end{eqnarray}

The evolution of the linear galaxy bias $b$ and the cross-correlation
coefficient $r$ is therefore given by:
\begin{eqnarray}
\label{bstochastic}
&& b = \left[ 1 + 2 {\langle \Delta_0 \delta_m \rangle \over \langle \delta_m^2 \rangle}
+ {\langle \Delta_0^2 \rangle \over \langle \delta_m^2 \rangle} \right]^{1/2} \\
\label{rstochastic}
&& r = {1\over b} \left[ 1 + {\langle \Delta_0 \delta_m \rangle \over \langle \delta_m^2 \rangle} \right] \;.
\end{eqnarray}

Recall that $\Delta_0$ is independent of time (Eq. [\ref{deltadiff1}]). 
Assuming the matter fluctuation $\delta_m$ grows, we can see that at late times:
\begin{eqnarray}
\label{latetime1}
{\langle \Delta_0^2 \rangle \over \langle \delta^2_m \rangle} \ll 
{\langle \Delta_0 \delta_m \rangle \over \langle \delta^2_m \rangle} \ll 1
\end{eqnarray}
and therefore at late times:
\begin{eqnarray}
\label{latetime2}
&& b \sim 1 + {\langle \Delta_0 \delta_m \rangle \over \langle \delta_m^2 \rangle} \\
\label{latetime3}
&& r \sim 1 - {1\over 2} {\langle \Delta_0^2 \rangle \over \langle \delta_m^2 \rangle}
+ {1\over 2} \left[ {\langle \Delta_0 \delta_m \rangle \over \langle \delta_m^2 \rangle} \right]^2
\end{eqnarray}
and so, $b \rightarrow 1$ and $r \rightarrow 1$, as expected, with $r$ tending
to unity at a faster rate than $b$, although,
as emphasized before, cosmic acceleration slows the approach to unity considerably.

Our remarks in \S \ref{linear}
concerning the possible development of a scale dependence to the large scale
galaxy bias apply to $b$ and $r$ here as well.
In other words, 
the quantities that control the evolution of $b$ and $r$,
$\langle \Delta_0 \delta_m \rangle / \langle \delta^2_m \rangle$ and
$\langle \Delta_0^2 \rangle / \langle \delta^2_m \rangle$, can develop
a scale dependence (on $R$, $\Delta x$ or $k$), due to 
possible scale dependent modifications
of the Poisson equation.

\section{Nonlinear Bias}
\label{nonlinear}

Fry \& Gazta\~{n}aga \cite{jimenrique} introduced a systematic expansion that
relates the galaxy and matter fluctuations:
\begin{eqnarray}
\label{gdelta}
\delta_g = b \delta_m + {b_2 \over 2} \delta_m^2 + ...
\end{eqnarray}
To find the evolution of $b_2$, we need to keep terms
up to second order in fluctuations. Assuming gradient flow i.e.
${\bf v_m} = {\bf \nabla} \phi^v_m$ and
${\bf v_g} = {\bf \nabla} \phi^v_g$, Eqs. (\ref{momconserve}) and
(\ref{momgconserve}) tell us that
\begin{eqnarray}
&& (\phi^v_g - \phi^v_m)' + {a'\over a} (\phi^v_g - \phi^v_m) \\ \nonumber
&& \quad + {1\over 2} {\bf \nabla}(\phi^v_g - \phi^v_m) \cdot {\bf \nabla} (\phi^v_g + \phi^v_m) = 0 \;.
\end{eqnarray}
To linear order, this equation implies what we already know from Eq. (\ref{vdiff1}),
i.e. $(\phi^v_g - \phi^v_m)^{(1)}$ decays like $1/a$, where the superscript $\,^{(1)}$ denotes
the order we are considering. Whether $(\phi^v_g - \phi^v_m)^{(2)}$ decays depends
on the behavior of $(\phi^v_g + \phi^v_m)^{(1)}$ which depends on how
the Poisson equation (\ref{poisson}) might be modified. 
It can be shown that $(\phi^v_g - \phi^v_m)^{(2)}$ decays away if 
$(\phi^v_g + \phi^v_m)^{(1)}$ grows slower than $a'$ \cite{DequalA}.
Henceforth, we will assume so, following \cite{jimfry} who essentially assumed
${\bf v_g} = {\bf v_m}$ from the beginning.

Equation (\ref{massconserve}) can be rewritten, up to second order, as
\begin{eqnarray}
\label{massconserve2}
&& - {\bf \nabla} \cdot {\bf v_m} ({\bf x}) = \delta'_m ({\bf x}) \\ \nonumber
&& \quad - \int {d^3 \tilde k\over (2\pi)^3}
{d^3 k \over (2\pi)^3}
e^{i \tilde {\bf k} \cdot {\bf x}}
{{\bf k} \cdot \tilde {\bf k} \over k^2} \delta'_m({\bf k}) \delta_m(\tilde {\bf k} - {\bf k})
\end{eqnarray}
where all time dependence is left implicit, and ${\bf k}$ and $\tilde{\bf k}$ denote wavevectors.
We have abused the notation a bit by using $\delta_m$ to stand for both the real space
and Fourier space counterparts, distinguishing them purely by their arguments.
A similar equation holds for the galaxies as well. 
Assuming the difference ${\bf v_g} - {\bf v_m}$ decays away to second order
(or is zero to begin with), we have therefore
\begin{eqnarray}
\label{deltadiff2order}
&& [\delta_g({\bf x}) - \delta_m({\bf x})]' = \int {d^3 \tilde k\over (2\pi)^3} {d^3 k \over (2\pi)^3}
e^{i \tilde {\bf k} \cdot {\bf x}}
{{\bf k} \cdot \tilde {\bf k} \over k^2} \\ \nonumber 
&& \quad [\delta'_g({\bf k}) \delta_g(\tilde {\bf k} - {\bf k}) - 
\delta'_m({\bf k}) \delta_m(\tilde {\bf k} - {\bf k})]
\end{eqnarray}
which holds to second order. Expanding $\delta_g$ as in 
Eq. (\ref{gdelta}), the first order terms in the above expression tell us
that $(b - 1)$ decays like $1/\delta_m^{(1)}$, consistent with the finding
in \S \ref{linear}. The second order terms go like
\begin{eqnarray}
\label{deltadiff2order2}
&& [(b-1)\delta_m^{(2)} + {b_2 \over 2} (\delta_m^{(1)})^2]'
= \\ \nonumber
&& \int {d^3 \tilde k\over (2\pi)^3} {d^3 k \over (2\pi)^3} 
e^{i \tilde {\bf k} \cdot {\bf x}}
{{\bf k} \cdot \tilde {\bf k} \over k^2} 
{\delta_m^{(1)}}' ({\bf k}) \Delta_0^{(1)} (\tilde {\bf k} - {\bf k})
%%{\delta_g^{(1)}}'({\bf k}) \delta_g^{(1)} (\tilde {\bf k} - {\bf k}) - 
%%{\delta_m^{(1)}}'({\bf k}) \delta_m^{(1)} (\tilde {\bf k} - {\bf k})]
\end{eqnarray}
where we have suppressed the ${\bf x}$ dependence on the left hand side, and
the time dependence on both sides. Here,
$\Delta_0^{(1)}$ is the Fourier counterpart of $\Delta_0$ in Eq. (\ref{linearbias}),
and is therefore a first order quantity, and denoted as such.
Note also that it is time independent.

It is straightforward to see that, to leading order:
\begin{eqnarray}
\label{b2evolve}
b_2 = O[\delta_m^{(2)}/{\delta_m^{(1)}}^3] +  O[1/\delta_m^{(1)}]
\end{eqnarray}
One generally expects that $\delta_m^{(2)} \sim {\delta_m^{(1)}}^2$, and therefore
$b_2$ tends towards zero at late times as long as $\delta_m^{(1)}$ grows.
The same caveat applies as before: cosmic acceleration significantly slows
the relaxation of $b_2$.

It is noteworthy that $b_2$ develops a scale dependence even if it started off
scale independent. This is true even if the Poisson equation (\ref{poisson}) holds \cite{jimfry}.
The same is not true for $b$: the Poisson equation (\ref{poisson}) guarantees
$b$ stays scale independent if it started off so, as noted in \S \ref{linear}.

\section{Discussion}
\label{discuss}

Our results on the evolution of bias are summarized by
Eq. (\ref{linearbias}), (\ref{linearbias2}) or (\ref{linearbias3}) for the linear bias,
by Eqs. (\ref{bstochastic}) and (\ref{rstochastic}) for the (linear) stochastic bias,
and by Eq. (\ref{b2evolve}) for the nonlinear bias.
The Poisson equation (\ref{poisson}) is not assumed in deriving these results, therefore
leaving open the possibility that gravity is modified or new sources of fluctuations (such
as clustered dark energy) exist.
The only assumptions made are mass and number conservation, and that matter
and galaxies are subject to the same force. For the linear bias and
the stochastic bias, it is not even necessary that the force be the gradient of
some potential. 

Following these assumptions,
our main conclusions are: (1) the bias tends to relax towards unity, but this
relaxation is significantly slowed by the diminishing structure growth once cosmic acceleration commences;
(2) the correlation coefficient $r$ tends towards unity faster than the bias $b$ does;
(3) the large scale (linear) bias dynamically develops a scale dependence if modifications to the
Poisson equation are scale dependent.

Conclusion (1) implies, everything being equal, galaxies that formed at
higher redshifts tend to have a more relaxed bias today. The stalling of bias relaxation
by cosmic acceleration, however, implies we cannot automatically conclude
that such galaxies have close to unity bias today, unless they had an initial bias
that was fairly close to unity.
In other words, one cannot freely approximate
$\delta_m \sim \delta_g$ by appealing to bias relaxation alone.
This is unfortunate, because $\delta_g$ is about the only observational handle
we have on $\delta_m$ without assuming the Poisson equation.
Other observations, such as the microwave background and gravitational lensing,
generally probe the metric perturbations, i.e. the left hand
side of the Poisson equation \cite{LHS}. A direct test of the Poisson equation requires
independent knowledge of $\delta_m$, i.e. the right hand side.
It appears one has to be content with indirect tests via the linear growth
factor, such as those widely discussed in the literature \cite{modgravref}.
They test for
how $\phi$ changes with time, which the system of equations (\ref{massconserve}), (\ref{momconserve})
and (\ref{poisson}) predicts. 
The relation between $\phi$ and $\delta_m$ is tested only
in so far as the prediction for linear growth
from this whole system of equations is tested. 
To go beyond this, one would need some independent method for constraining
the galaxy bias. Standard methods to do so carry additional baggage. 
Methods using higher moments, for instance, assume Gaussian initial conditions
as well as the Poisson equation itself.

Conclusion (3) is, in our view, the most interesting one.
It suggests a simple way to test for scale dependent modifications
to the Poisson equation, which 
are expected in many existing theories of modified
gravity or clustered dark energy (see \S \ref{linear}). 
One can measure the relative large scale (linear) bias of two different populations of galaxies which
are at different stages of bias relaxation (Eq. [\ref{relativebias}]).
The standard paradigm of general relativity $+$ uniform dark energy predicts
a scale independent linear bias (both at birth and in its
subsequent evolution). Checking for departures from scale independence 
therefore constitutes a test of the paradigm.
As illustrated in the top panels of Fig. \ref{bevolveB.yukawa} and \ref{bevolveB.DGP}, 
this test is best carried out by comparing overbiased and underbiased galaxies.
It is also important to focus on large scales, since the bias is expected to
develop scale dependence on nonlinear scales.

It was recently noted by \cite{nonGauss} that primordial non-Gaussianity could give rise to
a scale dependent large scale bias as well.
Distinguishing this possibility from 
modifications to the Poisson equation should be feasible since there are other
more direct ways to test for primordial non-Gaussianity, such as from the 
microwave background anisotropies \cite{paolo}.

As emphasized in \S \ref{intro}, our calculation of the bias evolution
assumes galaxy number conservation, just as in \cite{jimfry}.
Ultimately, one would like to go beyond the treatment in this paper, and
formulate a complete theory of halo bias accounting for formation, mergers and dynamics,
extending the excursion set theory \cite{mowhite,shethtormen} beyond
the standard (general relativity $+$ uniform dark energy) paradigm.
Such a theory would depend on the details of how the Poisson equation
is modified. 
%It would be enlightening to work this out in simple cases.
It is interesting to note that the standard excursion set theory predicts a bias evolution
that is consistent with the one derived here: $b(a) - 1 \propto 1/D(a)$
where $D$ is the linear growth factor (the halos
are identified at some scale factor $a_0 < a$).
This suggests our simple dynamical calculation already captures much
of the relevant physics. An additional bonus of extending the excursion
set theory beyond the standard paradigm is that one could address the issue
of whether galaxies/halos could be born with a scale dependent bias in
non-standard theories.
We hope to pursue these issues in the future.

%For this test, there is no need to carefully identify a population of bias-relaxed galaxies.
%All one needs is two sets of galaxies which one has reasons to believe are at different
%stages of relaxation. If scale dependent modifications to the Poisson equation exist,
%it would seem like a conspiracy indeed for any two sets of, say old and young, galaxies to
%share exactly the same scale dependent bias, making the relative bias scale independent.
%It is worth noting that primordial non-Gaussianity could give rise to
%a scale dependent large scale bias as well, as recently pointed out by \cite{nonGauss}. 
%Distinguishing this possibility from 
%modifications to the Poisson equation should be feasible since there are other
%more direct ways to test for primordial non-Gaussianity, such as from the 
%microwave background anisotropies \cite{paolo}.
%We should also point out that we have assumed a scale independent bias at
%the formation of galaxies. Any scale dependence at birth would likely make
%it even easier to carry out our proposed test.
%It is actually not unreasonable to expect a scale dependent bias at birth 
%in those theories of modified gravity or clustered dark energy which introduce new scales
%to the problem. It would be interesting to work out the predictions for simple models
%of this type.

\acknowledgments

We thank Enrique Gazta\~{n}aga, Roman Scoccimarro, Ravi Sheth, Sheng Wang and
Pengjie Zhang for discussions. LH thanks the Fermilab Theoretical Astrophysics Group for 
a reunion that stimulated some of these discussions.
Research for this work is supported by the DOE, grant DE-FG02-92-ER40699,
and the Initiatives in Science and Engineering Program
at Columbia University. 

%\appendix

%\section{Incorporating the Alcock-Paczynski Effect}
%\label{app:AP}

\newcommand\spr[3]{{\it Physics Reports} {\bf #1}, #2 (#3)}
\newcommand\sapj[3]{ {\it Astrophys. J.} {\bf #1}, #2 (#3) }
\newcommand\sapjs[3]{ {\it Astrophys. J. Suppl.} {\bf #1}, #2 (#3) }
\newcommand\sprd[3]{ {\it Phys. Rev. D} {\bf #1}, #2 (#3) }
\newcommand\sprl[3]{ {\it Phys. Rev. Letters} {\bf #1}, #2 (#3) }
\newcommand\np[3]{ {\it Nucl.~Phys. B} {\bf #1}, #2 (#3) }
\newcommand\smnras[3]{{\it Monthly Notices of Royal
        Astronomical Society} {\bf #1}, #2 (#3)}
\newcommand\splb[3]{{\it Physics Letters} {\bf B#1}, #2 (#3)}

\newcommand\AaA{Astron. \& Astrophys.~}
\newcommand\apjs{Astrophys. J. Suppl.}
\newcommand\aj{Astron. J.}
\newcommand\mnras{Mon. Not. R. Astron. Soc.~}
\newcommand\apjl{Astrophys. J. Lett.~}
\newcommand\etal{{~et al.~}}


\begin{thebibliography}{99}
\bibitem{frame} Physics should be frame-independent, but the
statement of whether matter is minimally coupled is frame-dependent.
For instance, in scalar tensor theories, matter that is minimally
coupled in the Jordan frame is not minimally coupled in the Einstein frame.
Here, we assume minimal coupling in the frame where gravity appears
modified (e.g. Jordan frame in the context of scalar tensor theories). 
Of course, if minimal coupling holds in the Einstein frame, one is
back to Einstein gravity.
\bibitem{wileynote} Specifically, we use $\Omega_m = 0.27$, $\Omega_{\rm de} = 0.73$
and $w=-0.6$ ($w_{\rm grow}$ in the sense of \cite{wiley}) to compute $\delta_m^{\rm Poisson}$.
\bibitem{DequalA} 
If $(\phi^v_g + \phi^v_m)^{(1)}$ 
grows as fast as the scale factor $a'$, it can be shown that
$(\phi^v_g - \phi^v_m)^{(2)}$ approaches a constant. It turns out this would not change
our conclusion on $b_2$, but would complicate the derivation a little bit.
In any case, the standard (general relativity $+$ uniform dark energy) paradigm
typically gives a growth to $(\phi^v_g + \phi^v_m)^{(1)}$ 
that is slower than $a'$, and modifications of gravity to
accommodate the observed cosmic acceleration generally gives a growth that is even slower
(because gravity is made weaker on large scales).
\end{thebibliography}
\end{document}